\documentclass[spanish,notitlepage,11pt,preprint]{article} 

\usepackage[ansinew]{inputenc} 
\usepackage[spanish]{babel}    

\usepackage{amsmath}
\usepackage{amsfonts}

\usepackage[colorlinks=true,urlcolor=blue,linkcolor=blue, citecolor=blue]{hyperref}

\usepackage{graphicx}
\usepackage{geometry}           
\usepackage{epstopdf}
\usepackage{fancyhdr} 

\title{Exposición Temprana de Nativos Digitales en Ambientes, Metodologías y Técnicas de Investigación en la Universidad}
\author{
\textbf{H. Asorey}\thanks{Corresponding author: \href{mailto://hasorey@uis.edu.co}{hasorey@uis.edu.co}}, \\ 
\textit{Escuela de Física, Facultad de Ciencias, }\\
\textit{Universidad Industrial de Santander,} \\ 
\textit{Bucaramanga 680002, Colombia; y} \\
\textit{Laboratorio de Detección de Partículas y Radiación } \\
\textit{Centro Atómico Bariloche e Instituto Balseiro (CNEA/UNC) } \\
\textit{San Carlos de Bariloche - Río Negro - Argentina;} \\
\textbf{L. A. Núñez} \\ 
\textit{Escuela de Física, Facultad de Ciencias, }\\
\textit{Universidad Industrial de Santander,} \\ 
\textit{Bucaramanga 680002, Colombia; y} \\
\textit{Departamento de Física, Facultad de Ciencias } \\
\textit{Universidad de Los Andes, Mérida 5101, Venezuela;} \\
\textbf{C. Sarmiento-Cano} \\ 
\textit{Escuela de Física, Facultad de Ciencias, }\\
\textit{Universidad Industrial de Santander,} \\ 
\textit{Bucaramanga 680002, Colombia} 
}
\date{\today}                                           

\begin{document}
\maketitle
\begin{abstract}

{\textbf{Resumen}}

Conscientes de la problemática relacionada con la motivación en el estudio de
las carreras científicas, presentamos dos experiencias en las cuales se expone
a estudiantes universitarios a ambientes, metodologías y técnicas del
descubrimiento abordando problemas contemporáneos. Estas experiencias se
desarrollaron en dos contextos complementarios: un curso de Introducción a la
Física, en el cual se motiva a estudiantes de Física a la participación
temprana en investigación, y un semillero de investigación multidisciplinario
para estudiantes avanzados de pregrado en Ciencias e Ingeniería, el cual
produjo incluso contribuciones a eventos internacionales. Si bien los
resultados son preliminares y requieren más ediciones para ser estadísticamente
significativos, consideramos que son alentadores. En ambos entornos se observó
un aumento de la motivación de los estudiantes para continuar sus carreras
haciendo énfasis en la investigación. En este trabajo, además de presentar el
marco contextual en el cual se soportan las experiencias, describimos seis
actividades concretas para vincular a los estudiantes con los ambientes de
descubrimiento, las cuales creemos pueden ser replicadas en ambientes similares
de otras instituciones educativas.\\

{\textbf{Abstract}}

Being aware of the motivation problems observed in many scientific oriented
careers, we present two experiences to expose to college students to
environments, methodologies and discovery techniques addressing contemporary
problems. This experiences are developed in two complementary contexts: an
Introductory Physics course, where we motivated to physics students to
participate in research activities, and a multidisciplinary hotbed of research
oriented to advanced undergraduate students of Science and Engineering (that
even produced three poster presentations in international conferences).
Although these are preliminary results and require additional editions to get
statistical significance, we consider they are encouraging results. On both
contexts we observe an increase in the students motivation to orient their
careers with emphasizing on research. In this work, besides the
contextualization support for these experiences, we describe six specific
activities to link our students to research areas, which we believe can be
replicated on similar environments in other educational institutions.\\

{\bf{Palabras claves:}} Educación en Física, Articulación
Docencia-Investigación, Herramientas TIC, Ciencia de Datos.

\end{abstract}

\section{Introducción}
A partir de los años 70 hubo un cambio en el modo de producción del sistema
capitalista en el cual pasamos de una sociedad industrial a una informacional.
En esta nueva era de la humanidad la información ha ido transformando a la
economía en el mismo sentido que la industria transformó la actividad agraria
en industrial. No es retórico que estamos en la sociedad del conocimiento y/o
de la información: información y conocimiento son insumos y, a la vez,
productos en esta nueva economía\,\cite{Castells2000,Castells2001}. Las
actividades científicas y tecnológicas no se escapan a este proceso y se han
convertido en e-actividades que difieren de lo que habíamos venido haciendo en
términos metodológicos, funcionales y, sobre todo, en la manera como nos
organizamos para crear y diseminar el conocimiento que producimos. 

Del mismo modo que fue necesario instruir a los pobladores de una sociedad
agraria para poder seguir instrucciones y operar las nacientes maquinarias de
la era industrial, hoy es indispensable involucrar a la mayor parte de la
sociedad en la producción de conocimiento. Cada vez más y con mayor frecuencia,
se desarrollan proyectos de investigación en los cuales se teje un complejo
entramado de relaciones entre profesionales y aficionados y resuenan vocablos
como \textit{crowdsourcing}\,\cite{Travis2008} y ciencia
ciudadana\,\cite{Gura2013,KybaEtal2013,Zastrow2014}. 

Esta necesidad de articular docencia e investigación, de exponer tempranamente
a los estudiantes a los ambientes de descubrimiento, viene siendo discutida en
varios escenarios\,\cite{AndersonEtal2011} y  consolidándose a través de una
variedad de propuestas pedagógicas con varios enfoques y perspectivas (ver por
ejemplo\,\cite{ChevalierBuckles2013, FraserEtal2014, Mazur2014} y referencias
allí citadas). 

Pero no es solo el impulso vital de la época, ese tácito consenso social que
estimula a las organizaciones actuales a producir información y conocimiento
más y más rápido, es la necesidad de motivar a las nuevas generaciones de
estudiantes, nativos digitales\cite{Prensky2001}, para quienes el
descubrimiento de como funcionan infinidad de dispositivos es parte de su
cotidianidad. Definitivamente, incorporar el descubrimiento en la práctica
pedagógica de los primeros cursos de ciencias es esencial en estos tiempos de
cambio.

La Física en el Siglo XXI es una disciplina en rápida evolución que demanda
constantemente nuevos desarrollos y técnicas de análisis que permitan abarcar
un creciente número de preguntas cada vez más complejas. Esta evolución y sus
nuevas formas de organización para producir conocimiento (global y colectiva)
necesitan ser ilustradas en los distintos cursos de la carrera desde las etapas
más tempranas de la formación de los futuros investigadores. Es en los primeros
semestres cuando los estudiantes deben vislumbrar los alcances de la Física
como disciplina de estudio del mundo que los rodea, y ensayar un mínimo de
métodos, herramientas y técnicas indispensables para la investigación en
ciencias.  
 
Si bien existe una bien documentada de lista de prácticas exitosas de
articulación docencia-investigación, la inmensa mayoría de los primeros cursos
en la enseñanza de la ciencia siguen siendo tradicionales: clases magistrales
con tiza y pizarrón\cite{FraserEtal2014}. Por ello queremos contribuir con
algunas experiencias que muestran estrategias para exponer a estudiantes de
educación superior a ambientes, metodologías y técnicas de investigación
abordando problemas contemporáneos. Las hemos realizado de forma que puedan ser
sostenidas en el tiempo, escalables a cursos de distintos tamaños y replicables
en otros ambientes de instituciones universitarias en nuestra América Latina.

Este trabajo fue organizado de la siguiente manera: en la próxima sección
analizamos el contexto de ese cambio de época y como incide en las nuevas
realidades de la enseñanza de las ciencias; en la sección \ref{Experiencias}
describimos dos experiencias de participación en investigación; seguidamente,
en la sección \ref{Resultados} mostramos algunos resultados muy preliminares de
estos ensayos; a continuación desarrollamos unas reflexiones generales sobre
ambas experiencias; finalmente, en los apéndices listamos las actividades
desarrolladas para generar esa exposición a la investigación científica. 

\section{El contexto de la producción temprana de conocimientos}\label{Contexto}

\subsection{La emergencia de la ciencia de datos y el aprendizaje colaborativo}

Los términos ``ciberinfraestructura'', ``e-ciencia'' y más recientemente uno
más amplio, ``e-investigación'', han sido acuñados para describir nuevas formas
de producción y diseminación del conocimiento
(ver\,\cite{HeyTrefethen2003,Foster2005,HeyTrefethen2005} y las referencias
allí citadas). Parte de los retos que habremos de enfrentar en esta nueva
manera de hacer ciencia son el de manejar, administrar, analizar y preservar un
``diluvio de datos''\,\cite{HeyTrefethen2003a}. Este alud de mediciones
convierte a los instrumentos en herramientas informáticas y la experimentación
no puede aislarse de términos como minería de datos. La avalancha de registros
de todo tipo viene generada por experimentos de escala mundial (aceleradores de
partículas, red de observatorios terrestres y satelitales e infinidad de los
más variados sensores dispersos geográficamente), los cuales desbordan toda
capacidad de manejo que no sea mediante uso intensivo de las TIC. 

Hoy en día, no es extraño encontrar colaboraciones que van de varias decenas
hasta miles de científicos, provenientes de cientos de instituciones radicadas
en decenas de países diferentes, trabajando mancomunadamente en un entorno muy
competitivo y a la vez, altamente colaborativo. Los estudiantes deben
prepararse desde las etapas más tempranas a trabajar en estos entornos. Desde
el aula, es posible fomentar el trabajo en equipo y colaborativo, mediante la
creación de grupos de trabajo formados por estudiantes escogidos al azar,
desalentando los cambios de personas entre grupos. De esta forma, se estimula a
los estudiantes para que se adapten y aprendan a trabajar en equipo con sus
compañeros de grupo, y eventualmente pulir asperezas para alcanzar el éxito en
las tareas encomendadas al grupo\,\cite{Scherr2009, Moliner2011}. Así, el
aprendizaje colaborativo ``parte de concebir a la educación como proceso de
socio-construcción que permite conocer las diferentes perspectivas para abordar
un determinado problema, desarrollar tolerancia en torno a la diversidad y
pericia para re-elaborar una alternativa conjunta''\,\cite{Calzadilla2002}.
Entonces, estos entornos se convierten en ``un lugar donde los estudiantes
deben trabajar juntos, ayudándose unos a otros, usando una variedad de
instrumentos y recursos informativos que permitan la búsqueda de los objetivos
de aprendizaje y actividades para la solución de
problemas''\,\cite{Wilson1995}.

Quizá no tengamos una conciencia clara de los profundos cambios que habrán de
experimentarse en nuestra actividad académica por esa necesidad de manejar y
analizar inmensos volúmenes de datos. Es tal la cantidad de información a la
cual hoy tienen acceso nuestros estudiantes, que debemos plantearnos una
reflexión en torno a los contenidos y a las metodologías que utilizamos
cotidianamente en la formación de estos futuros profesionales. Nuestra función
como docentes habrá de focalizarse en la enseñanza de los principios básicos en
ciencias y humanidades, proveyendo el adiestramiento necesario para que los
estudiantes puedan encontrar en la red la información pertinente y valorar su
calidad.  Si bien los ingentes volúmenes de datos provenientes de mediciones
reales y disponibles a través de la red, abren inmensas posibilidades para
hacer una docencia productora de nuevos conocimientos y, más aún, se comienzan
a ver los esfuerzos por utilizar estas herramientas y metodologías de la
e-investigación\,\cite{GraySzalay2002,BardeenEtal2006,Borgman2006} en la
educación, existe una resistencia bien marcada por parte de los mismos
investigadores en utilizar las TIC en su docencia
cotidiana\,\cite{Borgman2006,Wouters2006,FosterGibbons2005}. Nos aferramos a
los viejos paradigmas y visiones de la actividad científica. No percibimos, o
no queremos percibir, que nos cambió el panorama y nos resistimos a entrar en
la era informacional. 

\subsection{Nativos e inmigrantes digitales}

Las estadísticas de ingreso a nuestras universidades muestran que, en promedio,
más del $90\%$ de los estudiantes admitidos tienen menos de 20 años de edad
(ver para el caso de la Universidad Industrial de Santander
(UIS)\footnote{\url{http://www.uis.edu.co/planeacion/documentos/uisencifras/2013/index.html}}
página 58 de\,\cite{ArenasMonroy2013}) situándolos dentro de la generación de
los ``nativos digitales''\,\cite{Prensky2001}; crecidos en un entornos
computacionales, en los cuales el acceso  a consultas a
Wikipedia\footnote{\href{http://es.wikipedia.org}{wikipedia.org}} es casi
permanente y para quienes las numerosas aplicaciones para teléfonos celulares
ponen la suma del conocimiento humano al alcance de sus dedos. 

Estos nativos digitales, a lo largo de su vida, han dispuesto menos de $5000$
horas para lectura de libros, pero han estado más de $10000$ horas frente a
videojuegos, o $20000$ horas frente a la televisión\,\cite{Prensky2001}. Correo
electrónico, Google\footnote{\href{http://www.google.com/}{www.google.com}},
Facebook\footnote{\href{http://www.facebook.com/}{faceboook.com}},
Twitter\footnote{\href{http://twitter.com}{twitter.com}}, {\textit{blogging}},
son parte integral de sus vidas, donde el acceso a enormes volúmenes de
información en tiempo real ha cambiado radicalmente la forma en que los nativos
digitales procesan e interpretan los estímulos del mundo que los rodea. 

La facilidad para acceder a grandes volúmenes de información se refleja en el
aula: los estudiantes realizan consultas en tiempo real durante la clase, pero
no al docente, si no a la red, cuando el profesor utiliza un término que ellos
desconocen. Como docentes, debemos dejar de ser la única fuente de información
en el aula, para convertirnos en guías y mentores de un aprendizaje individual,
brindando una plataforma de conceptos y conocimientos básicos que los ayude a
evaluar la calidad de la información y discernir sobre la veracidad de una
ingente avalancha de datos e información que tienen frente a cada respuesta que
buscan en la red, usando indicadores como los que proponen, por ejemplo, 
Flanagin y Metzger\,\cite{FlanaginMetzger2007}. Al nivel institucional, la
Universidad debe comenzar a poner más énfasis en el aprendizaje (lo que hacen
los estudiantes) que en la docencia (lo que hacemos los profesores).

Los docentes, en la mayoría de los casos mayores de 30 años, somos
``inmigrantes digitales'', no nacimos pero nos vimos inmersos en un mundo nuevo
de rápida adaptación. Debemos ser capaces de lograr adaptarnos a este nuevo
entorno, captar la atención de los nativos digitales, hablar en su propio
lenguaje, introducir los conceptos a transmitir de la forma que ellos puedan
incorporar mejor: de lo estático a lo dinámico, de lo puntual a lo continuo, de
lo escrito a lo visual.

\subsection{Enseñanza y tecnologías de información y comunicación}

Es innegable la penetración que han tenido las TIC tanto en la investigación
como en la enseñanza. De las TIC se espera que, en resumen, logren mejorar la
adaptación al proceso enseñanza-aprendizaje. Sin embargo, las TIC, al ser
herramientas, no garantizan por si solas el cumplimiento de esos objetivos.
Depende de los docentes lograr que la implementación de TIC en el aula
faciliten y favorezcan el aprendizaje de los
estudiantes\,\cite{Selwyn2008,Selwyn2010}. Estudios recientes muestran que la
actitud del profesorado frente a las TIC es tanto o más importante que los
recursos TIC puestos a disposición de la práctica
educativa\,\cite{Tornero2013}. Es por ello que esta propuesta metodológica
utiliza herramientas de las TIC como forma de acercarse a los estudiantes,
nativos digitales en su amplia mayoría, al utilizar entornos que ellos dominan
y a los que están más acostumbrados. Esta estrategia justificaría por si sola
la necesidad de implementar TIC en el aula. Sin embargo, gran parte de estas
herramientas son las que los estudiantes deberán incorporar a lo largo de su
carrera tanto académica como científica: consultas a bases de datos en línea,
técnicas de análisis de datos y simulaciones físicas utilizando lenguajes
interpretados como python, o redacción de informes o trabajos científicos en
\LaTeX, mantener una discusión en un foro, o simplemente compartir información
en la nube.

\section{Experiencias de participación en investigación}

\label{Experiencias}

En esta sección presentamos dos experiencias de participación en investigación
con estudiantes de educación universitaria desarrolladas desde la Escuela de
Física de la Universidad Industrial de Santander, Bucaramanga-Colombia. 

La primera de éstas lo constituye el curso de Introducción a la Física para
estudiantes de nuevo ingreso a la carrera de Física y la segunda se refiere al
desarrollo de un semillero de investigación multidisciplinario, con estudiantes
entre sexto y octavo semestre de las carreras en Ciencias e Ingenierías. Ambas
iniciativas exponen a los estudiantes a metodologías y técnicas muy similares a
las que enfrentarán en su futuro profesional como investigadores, a saber:

\begin{itemize}
  \item temas de actualidad, con referencias bibliohemerográficas recientes;
  \item acceso a fuentes de datos reales disponibles en línea;
  \item manejo de herramientas computacionales para realizar el análisis de
    datos, simulaciones y posibilidad de contrastar resultados de experimentos
    con simulaciones y;
  \item redacción de un informe con formato de artículo con herramientas
    profesionales de composición de texto que, para el caso de la Física
    consideramos que es \LaTeX.
\end{itemize}

En el \ref{ApDocenciaInvestigacion} hacemos un listado detallado de las
actividades prácticas que articulan docencia-investigación y ciencia de datos.
Indicamos sus objetivos, las referencias y las acciones que deben cumplir los
estudiantes.

\subsection{Introducción a la Física para nativos digitales}

En la más reciente reforma del plan de estudios de la carrera de Física de la
Universidad Industrial de Santander (Bucaramanga-Colombia) se incluyó un curso
llamado Introducción a la Física, con un doble objetivo: nivelar en conceptos y
técnicas de Física a los estudiantes provenientes de distintos colegios y
motivarlos a continuar en la carrera de Física. 

Al ingresar a la Universidad, estos mismos estudiantes acarrean preconceptos,
en general negativos, respecto a la Física, basados en sus expectativas y
vivencias previas, y que se observan aún en aquellos que han elegido estudiar
esta disciplina. En el proceso de aprendizaje, estos preconceptos tienen una
influencia directa sobre la forma en la que los estudiantes se enfrentan a los
retos que implica comenzar (y porque no también cursar) esta
carrera\,\cite{RedishJefferySteinberg1998} y quizá inciden en el alto nivel de
deserción reportado en la carrera de Física de la UIS la cual, en el quinquenio
2008-2012\,\cite{ArenasMonroy2013}, tuvo un promedio de $66.2\%$ y con un
máximo de $75.6\%$ en el año 2010.

En el diseño curricular de nuestra propuesta se tuvo especial atención a la
condición de nativos digitales de los estudiantes, y por ello se abordó a
través de:
\begin{itemize}
  \item \textbf{Uso de bitácora y redes sociales como centro contenidos y
    discusión} La distribución de contenidos (presentaciones del curso, códigos
    Phyton, referencias de apoyo y vídeos relacionados) se concentraron en la
    bitácora
    (\textit{blog})\footnote{\url{http://halley.uis.edu.co/fisica_para_todos/}}.
    Se creó una cuenta de
    twitter\footnote{\url{https://twitter.com/fisicatodos}}, y un grupo abierto
    de
    Facebook\footnote{\url{https://www.facebook.com/groups/fisicareconocida/}},
    de acceso público. Los alumnos generaron en estas redes una dinámica
    mediante la publicación de preguntas, inquietudes, dudas, resultados de sus
    actividades, vídeos y noticias de los medios relacionadas con la temática
    del curso. Todo el material está licenciado bajo la licencia
    {\emph{Creative Commons Attribution-NonCommercial-ShareAlike 4.0
    International (CC BY-NC-SA 4.0)}}\footnote{Ver
    \url{http://creativecommons.org/licenses/by-nc-sa/4.0/} y
    \url{http://creativecommons.org/licenses/by-nc-sa/4.0/legalcode}}.
  \item \textbf{Acceso a la red en todo momento.} En tanto en las discusiones
    en el aula, como en las evaluaciones se incentivó el uso de fuentes de
    datos y conocimiento en línea como Wikipedia, google data, etc, para
    fundamentar las argumentaciones.
  \item \textbf{Capacitacion en el uso de herramientas computacionales} Tal y
    como se mencionó anteriormente, se utilizaron herramientas computacionales
    de uso cotidiano en ambientes de investigación, a saber: 
    \begin{itemize}
      \item
        Python\footnote{\url{http://www.python.org}}\,\cite{Gonzalez-Duque2010}
        es un lenguaje de programación de alto nivel, interpretado,
        multiparadigma y de propósitos generales, donde se pone especial
        énfasis en la simplicidad del código y su rápida implementación. Se
        caracteriza por su facilidad de uso y una rápida curva de aprendizaje,
        lo que lo hace ideal para su incorporación en el aula. Más aún, python
        ha sido adoptado por organizaciones mundiales de primer nivel, como
        Yahoo, Google, NASA, NWS y el
        CERN\footnote{\url{https://wiki.python.org/moin/OrganizationsUsingpython}},
        además de una enorme comunidad de usuarios en todo el mundo. Esto
        garantiza la existencia de una extensa base de ayuda en línea.
      \item
        Tracker\footnote{\url{http://https://www.cabrillo.edu/~dbrown/tracker/}}\,\cite{BrownCox2009}
        es un código java, gratuito, y con una interfaz muy simple de utilizar,
        que permite la edición, el análisis de vídeos y el modelado de
        situaciones físicas en un rango muy amplio de situaciones. Está
        específicamente diseñando para ser incorporado como herramienta para la
        educación en Física. Se incentivó a los estudiantes a realizar vídeos
        utilizando sus celulares sobre situaciones de su vida cotidiana para
        luego analizarlos utilizando Tracker.
      \item \LaTeX\,\cite{Lamport1994,MittelbachEtal2004} es un sistema de
        composición de textos orientado a la escritura de artículos, libros y
        tesis, de uso muy extendido en la comunidad científica. La calidad
        tipográfica del resultado final es comparable a la de imprentas
        profesionales. En las primeras clases de los laboratorios, se dará a
        los estudiantes un ejemplo de informe basado en \LaTeX para que lo
        desarrollen como base para las entregas que deben realizar a lo largo
        del curso. Se promocionó el uso de ambientes colaborativos en línea
        para la escritura de los informes, como por ejemplo
        {\emph{Write}}\LaTeX\footnote{\url{https://www.writelatex.com/}}.
    \end{itemize}
  \item \textbf{Desarrollo de Prácticas en Laboratorios Computacionales} La
    implementación de los laboratorios virtuales se apoyan en numerosas
    herramientas basadas en TIC, que en muchos casos requieren ser instaladas o
    apoyarse en recursos en línea, cuyo acceso es abierto pero en algunos casos
    son más limitados que sus versiones instaladas. Más aún, persiguiendo un
    criterio de equidad entre los estudiantes y la escalabilidad de esta
    propuesta, no es posible basar su aplicabilidad en el requisito de que los
    estudiantes dispongan de computadores portátiles que puedan utilizar
    durante las clases prácticas y también en sus casas.  Este criterio de
    equidad se logra montando el laboratorio virtual en aulas con computadoras
    disponibles en prácticamente toda Universidad. Con el fin de adaptarse a
    cualquier infraestructura disponible, se utilizan tecnologías de
    virtualización basadas en máquinas virtuales de sistema\,\cite{IBM2007},
    utilizando para ello la virtualización
    VirtualBox\footnote{\url{https://www.virtualbox.org/}} de Oracle, gratuita
    y disponible para todos los sistemas operativos. Con el fin de minimizar
    los recursos, la virtualización utiliza el sistema operativo Xubuntu 14.04
    para arquitecturas de 32 bits\footnote{\url{https://xubuntu.org/}}.
    Xubuntu, basado en la distribución
    Ubuntu\footnote{\url{https://www.ubuntu.com/}} es un sistema operativo de
    código abierto, gratuito y liviano, basado en el núcleo GNU/Linux, con una
    alto nivel de penetración tanto en los usuarios finales como en sistemas de
    computo de alto rendimiento, académicos y corporativos. La elección de este
    sistema garantiza la sostenibilidad en el tiempo y la escalabilidad a
    cursos con mayor número de estudiantes, al no depender de la compra de
    licencias de software, y además transmite a los estudiantes la filosofía
    del código abierto a los datos, a los sistemas y al conocimiento. La
    máquina virtual utilizada\footnote{Fuente disponible en
    \url{http://halley.uis.edu.co/archivos/xubuntu-f0-32.zip}.} está preparada
    para correr en una memoria USB de 16 GB, y tiene instalados todos los
    programas y aplicaciones necesarias para el normal desarrollo de las
    actividades del curso (python, gnuplot, tracker, etc). Esta forma de
    instalación resulta no invasiva para los sistemas de cómputo de la
    universidad, a la vez que garantiza la portatilidad de la máquina virtual
    que permite a los estudiantes continuar los desarrollos vistos en clases en
    sus hogares, o en cualquier otra computadora. En el
    \ref{ApDocenciaInvestigacion} se describen algunas de las prácticas
    desarrolladas en los laboratorios.
\end{itemize}

El curso supuso una participación de los estudiantes de seis horas por semana
en aula y cuatro de trabajo independiente. Las seis horas de participación en
aula se dividieron en tres encuentros semanales de dos horas cada uno:  
\begin{itemize}
  \item dos horas para la discusión del contenido programático, organizado en
    cuatro módulos: Introducción y Herramientas Matemáticas, Mecánica,
    Electricidad y Ondas (con una duración de 3, 6, 5 y 2 semanas
    respectivamente, 16 semanas en total); 
  \item treinta minutos de reforzamiento, sesenta para un ambiente SOLE (por
    sus siglas en inglés de {\textit{Self Organized Learning Environment}}, ver
    \ref{ApendSOLE}) y treinta de encuentro con la profesión a través de
    charlas con investigadores;
  \item dos horas de laboratorio para la capacitación en herramientas
    computacionales para la simulación, tratamiento de datos y elaboración de
    informes técnicos.
\end{itemize}

\subsection{Semillero de Ciencia de Datos}

En Colombia, la idea de semilleros de investigación surge como una iniciativa
para la formación de generaciones de relevo de investigadores y se remonta a la
década de los 80 y se consolida como un programa de alcance nacional con el
apoyo del Departamento Administrativo de Ciencias, Tecnología e Innovación
COLCIENCIAS para mediados de los años 90
(ver\cite{Molineros2008,QuinteroMunevarMunevar2008} y las referencias allí
citadas). Desde sus inicios esta idea adquiere variadas expresiones dependiendo
de la institución que los impulse. En la actualidad, esta idea se enfoca
principalmente a la vinculación de estudiantes avanzados en las actividades de
los grupos de investigación consolidados.

En nuestro caso esta iniciativa vinculó a estudiantes de pregrado de varias
disciplinas de la UIS con el estudio y descubrimiento de fenómenos astrofísicos
de altas energías y destellos de rayos gamma, mediante la minería y el análisis
de grandes volúmenes de datos. Para ello, se trabajó en el marco del proyecto
LAGO ({\emph{Latin American Giant Observatory}}), una colaboración de 10 países
hispanoamericanos (9 latinoamericanos y España) que estudian al Universo
Extremo, fenómenos de Meteorología y Climatología Espacial, y Radiación
Atmosférica con detectores de radiación cósmica
terrestres\,\cite{AllardEtal2008,Asorey2013}. Algunos detalles adicionales de
esta colaboración se describen en el \ref{LAGO}.

El semillero se desarrolló durante el año 2014 con 12 estudiantes de entre el
sexto y el octavo semestre\footnote{Todas las carreras tienen una duración de
10 semestres con una carga horaria típica de 14 a 16 horas de clases semanales}
de la carreras de Física, Ingeniería de Sistemas e Ingeniería Eléctrica y
Electrónica de la Universidad Industrial de Santander, Bucaramanga-Colombia. A
lo largo de 16 encuentros de cuatro horas por semana se logró
\begin{itemize}
  \item capacitar estudiantes en los conceptos básicos asociados al estudio de
    Astrofísica de altas energías, destellos de rayos gamma, minería y análisis
    de datos;
  \item preparar a los integrantes del semillero en el desarrollo de algoritmos
    computacionales para la búsqueda y clasificación de datos, enfocados en el
    estudio de fuentes de altas energías. Se hizo énfasis en la utilización de
    esquemas como el \textit{Moving Windows Average} para la búsqueda de
    destellos de rayos gamma y objetos astrofísicos de emisión aperiódica;
  \item producir un algoritmo de identificación automática y corrección de la
    línea de base de los detectores del proyecto LAGO (ver
    \ref{ApDocenciaInvestigacion}). El algoritmo desarrollado durante este
    curso ha sido aceptado por la colaboración LAGO como parte del protocolo
    base de análisis de datos oficial del proyecto\,\cite{Suarez2014}; y
  \item estudiar protocolos de preservación de datos a través de la Red de
    Repositorios LAGO, facilitando así el análisis de datos para el estudio y
    descubrimiento colectivo de fenómenos astrofísicos de altas energías en el
    marco del proyecto LAGO.
\end{itemize}

La metodología utilizada combinó exposiciones de los coordinadores del
semillero y de los propios estudiantes para afianzar las lecturas realizadas.
Además, se trabajó con frecuencia con el detector de agua Cherenkov -ubicado en
las instalaciones de la Universidad- para ejemplificar los procesos físicos de
las partículas al contacto con el agua, como son detectadas por el
fotomultiplicador, el tipo de señal que genera y los conteos que registra la
electrónica.

Para cumplir con los objetivos antes señalados se realizaron las siguientes
actividades:

\paragraph{Conceptos de Astropartícula} Durante 6 sesiones y con una frecuencia
quincenal  se realizaron seminarios enfocados al aprendizaje de los conceptos
básicos de la Física de astropartículas. Revisando los mecanismos de transporte
de rayos cósmicos, tales como Fermi de primer y segundo orden y la interacción
entre éstos y el campo magnético terrestre. Asimismo, se discutieron algunos
aspectos la Física de la medición del flujo de estas partículas con detectores
de agua Cherenkov de la colaboración LAGO.

\paragraph{Técnicas y herramientas de ciencia de datos} Para continuar con la
formación profesional de los miembros del semillero se les capacitó en el
desarrollo de algoritmos computacionales centrados en el tratamiento de grandes
volúmenes de datos. Se prestó especial atención a la utilización de los datos
disponibles por la colaboración LAGO en su repositorio LAGOData y al uso
arreglos de equipos computacionales de alto rendimiento.  Esta actividad se
distribuyó a lo largo en 10 sesiones, también con periodicidad quincenal y
complementaron la formación práctica de los seminarios de formación en Física
de Astropartículas. 

\paragraph{Minería en LAGODatos, búsqueda de GRB} En esta etapa, se discutieron
nuevos algoritmos en el análisis de los datos utilizados actualmente por la
Colaboración LAGO. La base de estos algoritmos consiste en re-escalar los
datos, de tal forma que permiten la búsqueda de destellos gamma, GRB y fuentes
de emisión periódica. Algunos de los resultados preliminares de esta actividad
fueron discutidos en los seminarios GIRG y en los encuentros virtuales de
análisis de la Colaboración LAGO. De esta forma se fortaleció la interacción
con estudiantes, profesores e investigadores del grupo de investigación y de
otros países miembros de la colaboración LAGO.

\paragraph{Socialización en línea} El desarrollo y los resultados de las
actividades realizadas en el semillero de investigadores fueron documentados en
un portal
web\footnote{\href{http://halley.uis.edu.co/SemilleroDatos/}{http://halley.uis.edu.co/SemilleroDatos/}}
destinado para ello así como también mediante herramientas WEB2.0. En este
portal se encuentran las presentaciones utilizadas durante las sesiones de
entrenamiento así como los códigos utilizados en las prácticas y aquellos
realizados por los estudiantes. Esto permitió la  generación de una bitácora de
consulta que sirve como herramienta divulgativa de las actividades realizadas
en el semillero. 

\section{Resultados Preliminares}\label{Resultados}

Para el caso de la asignatura Introducción a la Física, los cambios propuestos
se pusieron en práctica a lo largo de las ediciones 2013 y 2014 del curso. Tal
como se dijo, este curso corresponde al primer curso de Física de los
estudiantes que ingresan a la carrera de Física de la UIS. Para el año 2014,
luego de la evaluación de los primeros resultados en el año 2013, se
profundizaron aquellos factores que resultaron positivos y se corrigieron
aquellos en los cuales se observó que los estudiantes tuvieron mayores
inconvenientes. El proceso de enseñanza-aprendizaje se logró mediante la
evaluación continua y el permanente diálogo con los estudiantes durante el
curso. La identificación de algunas de las falencias del curso se obtuvo
gracias a la realización de encuestas anónimas al final de cada curso, cuyos
resultados se resumen a continuación. 

\begin{figure}[!ht]
  \begin{center}
    \includegraphics[width=0.45\textwidth]{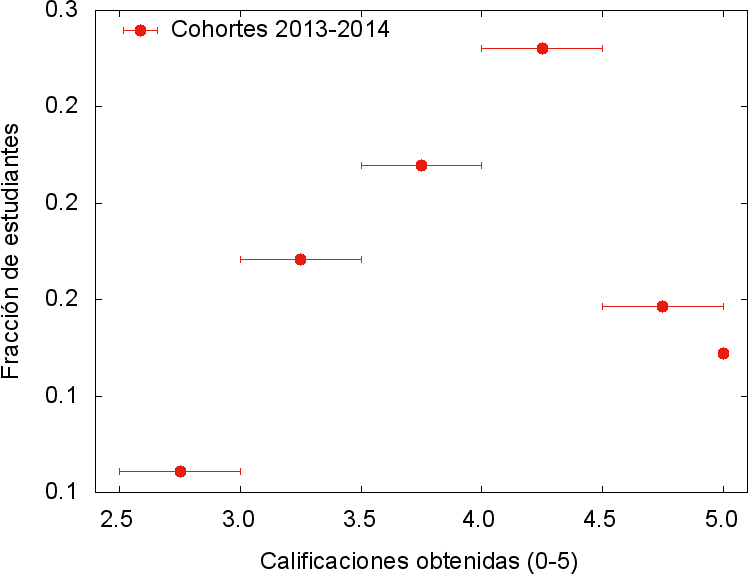}~~~\includegraphics[width=0.45\textwidth]{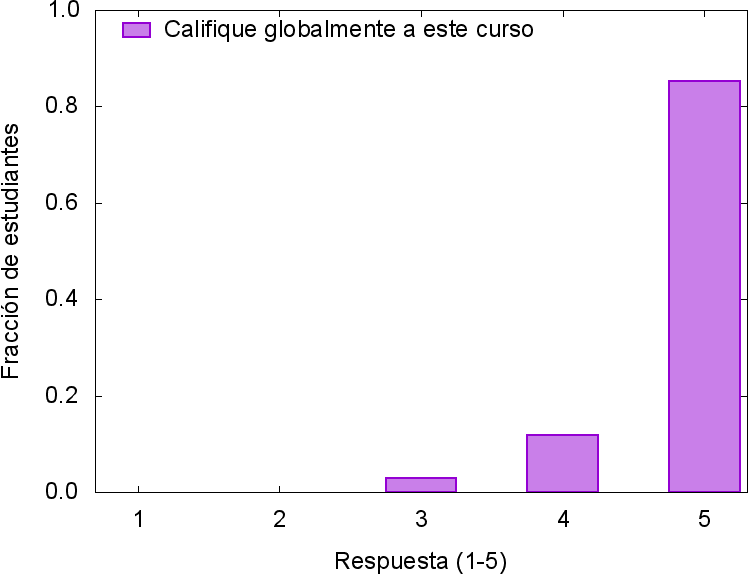}
  \end{center}
  \caption{Izquierda: Calificaciones finales (ver nota al pie \ref{calif} para
    el significado de las mismas) de los estudiantes que finalizaron el curso
    ($82/95=86.3\%$ de los estudiantes inscriptos). En total, $77/82=93.9\%$ de
    los estudiantes que finalizaron el curso fueron aprobados, $10/82=12.2\%$
    de ellos con la máxima calificación. Derecha: Respuestas dadas por los
    estudiantes durante las encuestas finales del curso ante la consigna
    ``Califique globalmente a este curso'' (véase la tabla \ref{taresultados}).
  }
  \label{notas}
\end{figure}

En términos generales, nuestra muestra se compone con 48 estudiantes de la
cohorte 2013 y 47 estudiantes de la cohorte 2014. Las calificaciones del curso
se conformaban mediante un promedio pesado entre las entregas de trabajos
prácticos, informes de laboratorios y exposiciones en formato charla que los
estudiantes debían presentar. Respecto al rendimiento de los estudiantes
durante el curso, se observó una gran disminución en la tasa de deserción
observada en esta asignatura respecto a años anteriores: $8/48=16.6\%$ para el
año 2013 y $5/47=10.6\%$ para el 2014\footnote{Debe aclararse que la deserción,
y especialmente en los primeros cursos, puede deberse a factores externos a un
curso en particular, como ser: el rendimiento en otros cursos, problemas de
índole personal, falta de adaptación a la vida universitaria, etc.}. De los
$82$ estudiantes que finalizaron la cursada, un total de $67$ aprobaron el
mismo y $10$ de ellos alcanzaron la máxima calificación, con un promedio de
$3.95$ y una desviación estándar de $0.7$\footnote{\label{calif} El rango de
calificaciones va de cero a cinco, con los siguientes significados: $0$,
reprobado; $(0-3)$, no aprobado; $[3-5)$, aprobado; y $5$ significa
sobresaliente.}. En el panel izquierdo de la figura \ref{notas} se muestra un
histograma de las calificaciones obtenidas por aquellos estudiantes que
finalizaron el curso. 

\begin{table}
\begin{tabular}[!hht]{|p{0.20\textwidth}|p{0.55\textwidth}|c|c|}
	\hline
	Consigna & Descripción & Promedio & Desvío \\
	\hline
  Califique globalmente a este curso & Apunta a que los estudiantes, utilizando
  sus experiencias y vivencias previas, califiquen esta nueva propuesta
  metodológica (ver histograma en el panel derecho de la figura \ref{notas}).&
  4.82 & 0.46 \\
	\hline
  No tuve dificultades para entender y aprovechar el curso & Se complementó con
  una pregunta descriptiva. $1$ significa que tuvo severas dificultades, $5$
  que no tuvo inconvenientes (ver figura \ref{resultados}). & 4.26 & 0.71 \\
	\hline
  Después de concluir el curso, me siento más interesado/a por estudiar Física
  & $1$ significa que está completamente en desacuerdo con la consigna, $5$ que
  está totalmente de acuerdo y el interés aumentó (ver figura
  \ref{resultados}). & 4.91 & 0.38 \\
	\hline
  Las charlas invitadas, ¿le sirvieron para tener una visión más global de la
  investigación en física? & Pregunta orientada a evaluar la experiencia de las
  charlas invitadas de profesores de la Escuela de Física (ver figura
  \ref{resultados}). & 4.85 & 0.50 \\
	\hline
  Me gustó la iniciativa del blog y disponer de los materiales en línea &
  Evaluación de la novedosa implementación de un blog con contenidos asociados
  al curso (ver figura \ref{resultados}). & 4.88 & 0.54 \\
	\hline
  Hubiera preferido que todas las entregas fueran individuales & Durante el
  curso se incentivó el trabajo colaborativo. Esta pregunta se orienta a
  evaluar la experiencia en ese proceso. El $65\%$ de los estudiantes prefirió
  la modalidad grupal. & N/A & N/A \\ 
	\hline
\end{tabular}
\caption{
Encuesta final de cursada para las cohortes 2013-2014 del curso de Introducción
a la Física de la Universidad Industrial de Santander. Las respuestas numéricas
se evaluaban entre $1$ (``{\textit{estoy completamente en desacuerdo con la
consigna''}}) y $5$ (``{\textit{estoy totalmente de acuerdo con la
consigna}}'').\label{taresultados}
}
\end{table}

Las encuestas consistieron en una serie de 30 preguntas. Las preguntas se
dividieron en ocho secciones, orientadas a evaluar distintas facetas: el curso,
los profesores, la relación con el curso, su experiencia de aprendizaje, su
preparación previa, la forma de evaluar e informaciones adicionales, tendientes
a contextualizar la situación personal de los estudiantes frente al curso, a la
Universidad y a su preparación previa para la vida universitaria. Muchas de las
preguntas aceptaban respuestas numéricas en el rango $[1-5]$ (siendo $1$ una
respuesta totalmente negativa o desfavorable, y $5$ una respuesta totalmente
positiva o favorable), mientras que otras aceptaban respuestas textuales para
recabar las opiniones personales de los estudiantes. Si bien las preguntas
fueron anónimas, para preservar la privacidad de nuestros estudiantes sólo
incluiremos en este trabajo aquellas respuestas que se orientan a su evaluación
de esta propuesta metodológica. Los resultados se muestran en la tabla
\ref{taresultados}.

\begin{figure}[!hh]
  \begin{center}
    \includegraphics[width=0.45\textwidth]{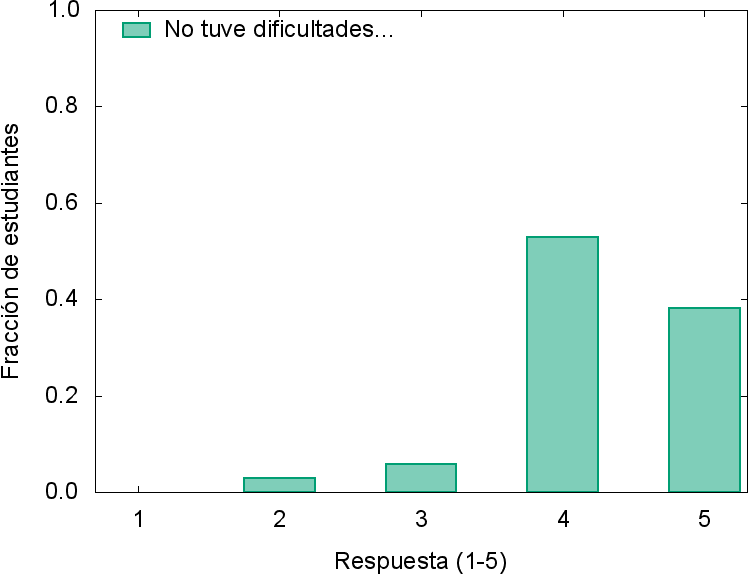}~
    \includegraphics[width=0.45\textwidth]{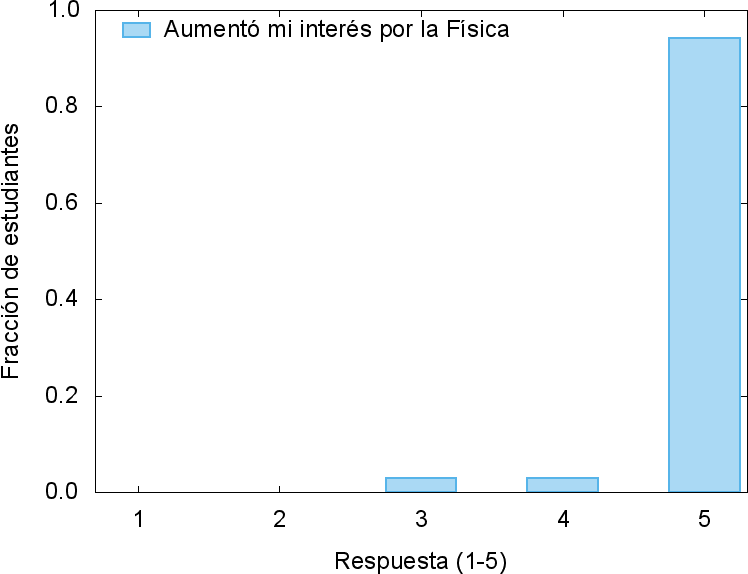}\\
    \includegraphics[width=0.45\textwidth]{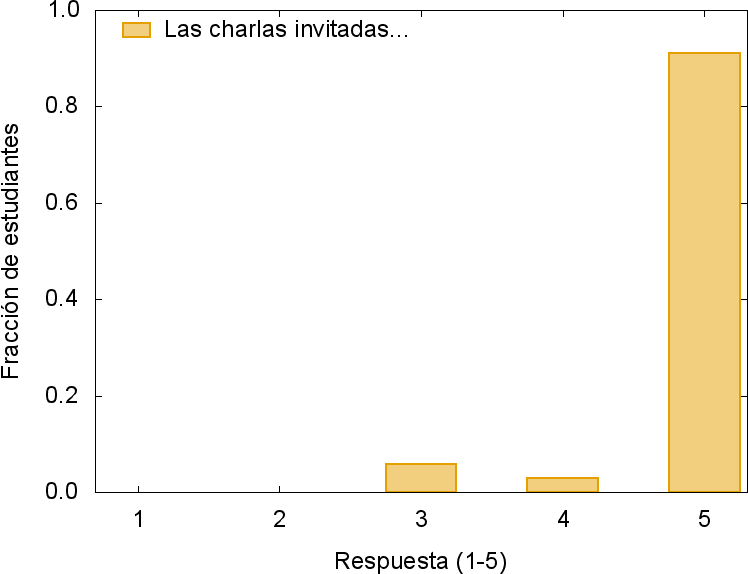}~
    \includegraphics[width=0.45\textwidth]{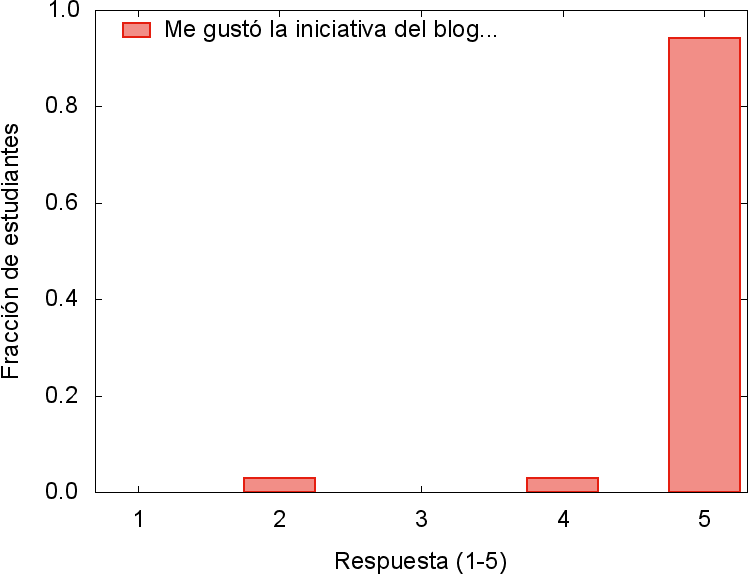}
  \end{center}
  \caption{Resultado de las encuestas finales de las cohortes 2013-2014. De
  izquierda a derecha y de arriba hacia abajo: ``No tuve dificultades para
  entender y aprovechar el curso'', ``después de concluir el curso, me siento más
  interesado/a por estudiar Física'', ``las charlas invitadas, ¿le sirvieron para
  tener una visión más global de la investigación en física?'', y ``me gustó la
  iniciativa del blog y disponer de los materiales en línea''. Ver explicaciones
  adicionales en el texto principal y en la tabla \ref{taresultados}.  }
  \label{resultados}
\end{figure}

Entre los resultados más relevantes del semillero podemos destacar: 

\paragraph{Dominio de herramientas y técnicas computacionales de alto
rendimiento para el manejo de grandes volúmenes de datos.} Los participantes
adquirieron destrezas en la utilización del sistema operativo LINUX y
programación en lenguaje Python. Dato la diversidad de las procedencia de los
participantes, hubo que nivelar algunos estudiantes quienes no estaban
familiarizados con esta plataforma de software libre. Se eligió
\textit{python} para realizar los cálculos y el análisis de datos fue por su
gran versatilidad, facilidad para aprenderlo y su robustez. Para esta
capacitación nos apoyamos en la plataforma en línea
\textit{codecademy}\footnote{\href{www.codecademy.com}{www.codecademy.com}} que
brinda un curso interactivo y general sobre este y otros lenguajes de
programación modernos. Para las prácticas se utilizaron los datos del proyecto
LAGO con los cuales se hicieron análisis básicos de cálculo de promedios y
desviación cuadrática. En esta fase se aprovechó para hacer una descripción
detallada de los datos y hacer pruebas con los datos de forma remota usando la
terminal de Linux.

\paragraph{Mejora del Algoritmo para identificar fuentes aperiódicas y creación
de código de validación.} Se estudió y mejoró el algoritmo utilizado para la
búsqueda de fuentes de astropartículas de emisión aperiódica. Teniendo en
cuenta que el tipo de fuentes que se buscan son destellos gamma se duplicó la
ventana de muestreo del algoritmo de la ventana móvil, de uno a dos minutos,
con lo que se logró aumentar el número candidatos. Asimismo, se desarrolló un
código\footnote{Códigos disponibles en: 
\url{http://halley.uis.edu.co/SemilleroDatos/codigos}} que revisa la
estabilidad de la línea base del detector generando una alarma de error cuando
las fallas se vuelven persistentes, este código está ahora en la fase de evaluación y
mejoramiento por parte de la Colaboración LAGO para ser incorporado al sistema
electrónico de adquisición de los detectores Cherenkov en agua del proyecto.

\paragraph{Incorporación de los estudiantes a grupos de investigación y
participación en congreso internacional.} Cumpliendo con los objetivos del
programa de semilleros, varios de estos estudiantes ingresaron al grupo de
investigación y realizan sus trabajos de grado, y algunos de los resultados han
sido presentados en el Simposio Latinoamericano de Física de Altas Energías (X
SILAFAE\footnote{\url{http://csi.uan.edu.co:82/conferenceDisplay.py?confId=1925}})\,\cite{Calderon2014,
Estupinan2014, Pinilla2014}.

\section{Reflexiones finales}\label{Reflexiones}

En este trabajo hemos descrito dos experiencias en las cuales se expone de
manera temprana a estudiantes que inician sus estudios universitarios a los
ambientes y metodologías actuales de producción de conocimiento. Estas
experiencias se han desarrollado en dos contextos complementarios:
\begin{itemize}
  \item uno formal, representado por un curso de Introducción a la Física para
    estudiantes de nuevo ingreso de la Carrera de Física de la Universidad
    Industrial de Santander en Bucaramanga-Colombia; 
  \item y otro informal desarrollado a través de un semillero de investigación
    de Ciencia de Datos para estudiantes avanzados de Física e Ingeniería,
    también de la Universidad Industrial de Santander.
\end{itemize}
Ambas experiencias resultaron exitosas, motivando a los estudiantes a incorporarse actividades de investigación con una perspectiva multidisciplinaria. Los estudiantes del nuevo ingreso de Física ratificaron su interés en continuar la carrera con una visión más clara de su futura vida profesional. Para el caso del semillero los participantes tanto de Ingeniería como de Física lograron generar contribuciones modestas pero originales en la mejora de algoritmos de manejo de grandes volúmenes de datos. Tres de estos aportes fueron aceptado como contribuciones a uno de los congresos regionales más importantes de la disciplina. 

En los últimos años se ha puesto especial atención a la relación entre el proceso de aprendizaje y la motivación de los estudiantes, con tres componentes reconocidos:
\begin{itemize}
  \item estados motivacionales, 
  \item ambientes de aprendizaje, y
  \item indicadores sociales de aprendizaje.
\end{itemize}

Tal como se explica en\,\cite{Fischer1997}, la actitud del docente frente a la
clase tiene influencias positivas o negativas en esas tres componentes, que
pueden afectar tanto positiva como negativamente al proceso de
enseñanza-aprendizaje. Desarrollar un entorno abierto, donde el flujo de
información no tiene un único sentido, donde el docente se convierte en un
observador y un consejero en lugar de un instructor. De esta manera, bajo la
guía del docente, el estudiante puede verificar la verdad o falsedad de sus
propias ideas o afirmaciones y así construir y mejorar sus propias ideas y
teorías. Esta actitud ayuda, además, a aumentar el nivel de auto-confianza del
estudiante e incrementa la interacción con el docente y con el resto de la
clase. Estudios recientes muestran también que la epistemología de los
estudiantes, es decir, sus ideas acerca del conocimiento y del aprendizaje,
afectarán la forma en que ellos aprenden\,\cite{Lising2005}. Es por ello, que
no sólo debemos enfocarnos en los contenidos programáticos si no también
escuchar a los estudiantes y entender que es lo que ellos piensan sobre la
física, e incorporar esas preconcepciones en la forma de encarar el curso.

Estamos conscientes que estos resultados son aún preliminares en función de la
limitada estadística disponible hasta el momento, pero son alentadores y
creemos que es imperioso insistir, por esta vía, a incorporar a los estudiantes
(y a la sociedad en general) a la producción de conocimientos, apoyando todos
los esfuerzos e iniciativas de ciencia ciudadana.  

Las clases de Física deben dejar de ser entornos silenciosos, debe haber una
activa y abierta discusión sobre los conceptos físicos que se están
desarrollando. Este entorno genera que los estudiantes se sientan actores y
partes, satisfaciendo su sentimiento de competencia y mejorando su interés por
la materia de estudio. La Física deja de ser un tema cerrado y definido, donde
la única verdad es la que dicen los libros y que es transmitida por el docente,
y se convierte en un tópico abierto y en evolución constante, donde el único y
crucial punto de referencia se da por el contraste de los modelos o esquemas
mentales con los resultados de los experimentos.

\section*{Agradecimientos}

Los autores reconocen el inestimable apoyo de la Vicerrectoría de Investigación
y Extensión de la Universidad Industrial de Santander y del Departamento
Administrativo de Ciencias, Tecnología e Innovación COLCIENCIAS bajo el
programa de Semilleros de Investigación en la Convocatoria No.617 año 2013.
Igualmente agradecen el apoyo financiero del Fondo Regional para la Innovación
Digital en América Latina y el Caribe, subvención 2013-314, que permitió
desarrollar el módulo de cambio climático  (\ref{CambioClimatico}) y gran parte
de los apoyos computacionales para ser descargados. Finalmente, dos de nosotros
(HA y CS) agradecemos el soporte financiero del proyecto de Articulación
Docencia-Investigación VIE-5751/2014. 

\bibliographystyle{unsrt}
\bibliography{asorey-nunez-sarmiento-introduccion-fisica}

\appendix
\renewcommand\thesection{\appendixname\ \Alph{section}}
\pagestyle{empty}

\section{Reforzamiento SOLE}
\label{ApendSOLE}
El ambiente SOLE (por sus siglas en inglés {\textit{Self Organized Learning
Environment}}, \,\cite{Mitra2010})  consiste en una actividad colectiva de
discusión en la cual los estudiantes se dividen en grupos de dos o tres
integrantes para trabajar mancomunadamente en actividades propuestas por ellos
mismos o por el docente. Durante la misma actividad cada grupo debe presentar
sus resultados y luego, junto con el docente, se obtienen las conclusiones
finales de la actividad. 

Cada hora SOLE se programó de la siguiente manera:
\begin{itemize}
\item  5 minutos para el planteo de la pregunta o actividad a realizar (docente
  o estudiante)
\item 35 minutos para el desarrollo del tema planteado (grupos de estudiantes)
\item 15 minutos para la exposición de las conclusiones de cada grupo
  (representante de cada grupo)
\item  5 minutos para desarrollar las conclusiones y el sumario (docente).
\end{itemize}

\section{El Proyecto LAGO}\label{LAGO}

El propósito del proyecto LAGO es el diseño, la construcción, la puesta en
marcha y la operación del Observatorio Gigante Latinoamericano (de allí sus
siglas en inglés LAGO: {\emph{Latin American Giant Observatory}}), un
observatorio extendido de astropartículas a escala
global\footnote{\href{http://lagoproject.org}{lagoproject.org}}). Sus
principales objetivos científicos se orientan a la investigación básica en
Astropartículas en tres líneas: el Universo Extremo, fenómenos de Meteorología
y Climatología Espacial, y Radiación Atmosférica con detectores de radiación
cósmica terrestres\,\cite{AllardEtal2008,Asorey2013}.

\begin{figure}
  \begin{center}
    \includegraphics[width=0.6\textwidth]{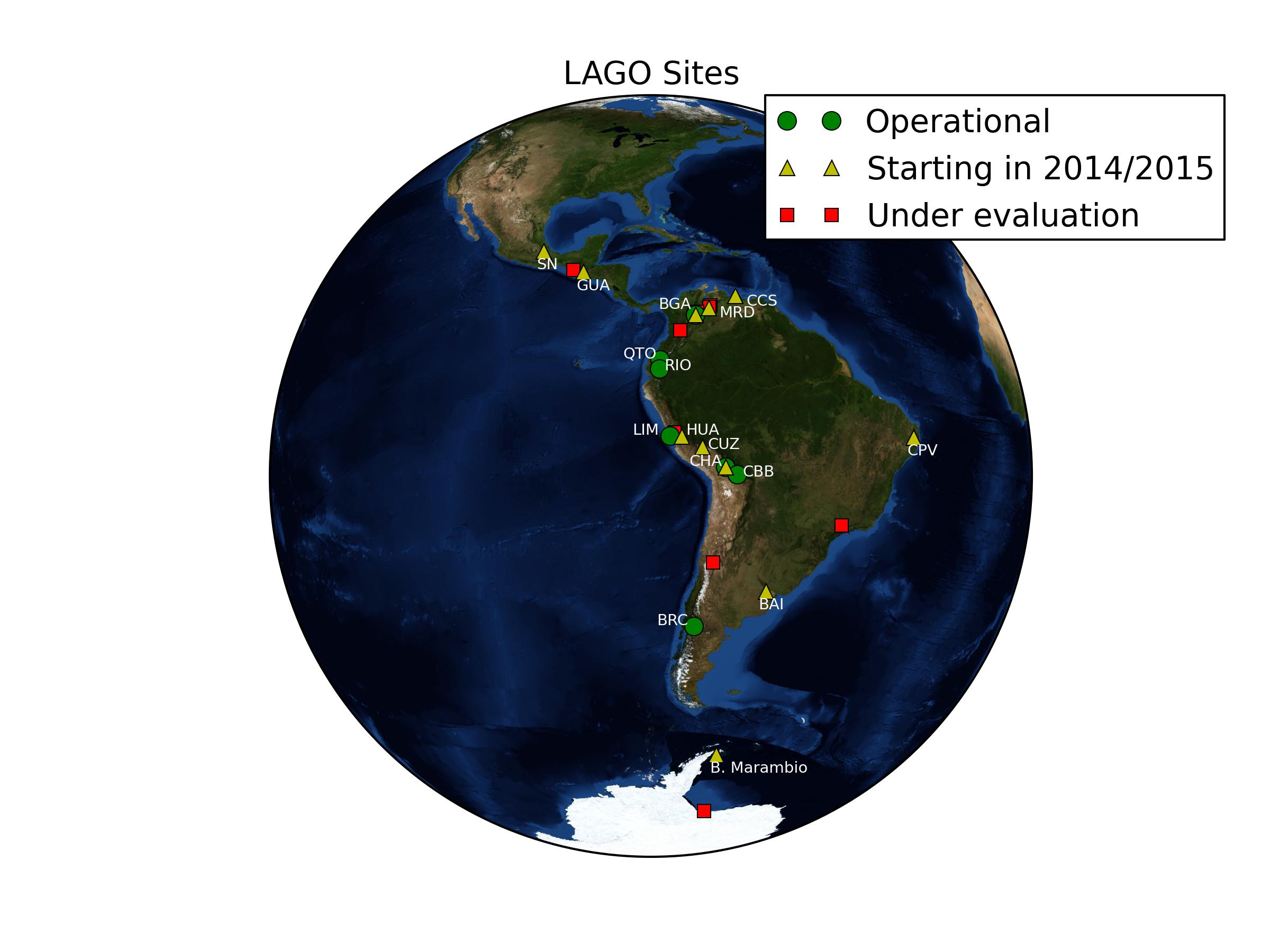}
    \caption{
      Distribución de los detectores LAGO en toda América Latina. Se muestran
      en verde los detectores operativos, en amarillo los que entran en
      funcionamiento en el bienio 2014-2015 y en rojo aquellos lugares
      actualmente en estudio para la instalación de nuevos detectores.
    }
    \label{LAGOmapa}
  \end{center}
\end{figure}

El proyecto LAGO es operado por la Colaboración LAGO, una red no centralizada,
distribuida y colaborativa, integrada por investigadores y estudiantes de
varias instituciones de los diferentes países miembros del proyecto LAGO:
Argentina, Bolivia, Brasil, Colombia, Ecuador, España, Guatemala, México, Perú
y Venezuela. En la figura \ref{LAGOmapa} se muestra la ubicación de las
instalaciones LAGO y algunas de los posibles sitios de instalación de futuros
sitios de la red de detectores del proyecto.  La colaboración LAGO también
mantiene una estrecha cooperación con investigadores europeos del Institut
National de Physique Nucléaire et de Physique des  Particules (IN2P3) de
Francia, de la Universidad de Granada, en España y el Istituto Nazionale di
Fisica Nucleare (INFN) en Italia.

LAGO dispone todas las características para desarrollar un espacio de
apropiación tecnológica, una comunidad virtual multidisciplinaria,
geográficamente distribuida, que coopera en torno a un proyecto de e-Astronomía
en América Latina, utilizando infraestructura de redes avanzadas.

\section{Docencia-investigación en Ciencia de Datos}
\label{ApDocenciaInvestigacion}

\subsection{Exoplanetas y leyes de Kepler}

Los exoplanetas, también llamados planetas extrasolares, son planetas que
orbitan en torno a estrellas diferentes el Sol, conformando sistemas
planetarios con características distintivas a nuestro Sistema Solar. Los
primeros exoplanetas fueron descubiertos en la década de los 90, con el
descubrimiento de tres objetos sub-estelares orbitando al púlsar PSR
B1257+12\cite{Wolszczan1992}, y el planeta 51 Peg b, descubierto en 1995, en
órbita en torno a la estrella 51 Peg en la constelación del
Pegaso\cite{Mayor1995}. Hoy en día, hay más de mil sistemas planetarios con
casi dos mil exoplanetas confirmados, con otros tres mil exoplanetas candidatos
descubiertos por la sonda Kepler\footnote{\url{http://kepler.nasa.gov/}}. El
bien conocido interés de los estudiantes de Física y Ciencias en general por la
astronomía puede canalizarse entonces para la introducción de conceptos
estadísticos, como la media o el desvío, o las leyes del movimiento planetario
usando datos reales de exoplanetas disponibles en línea, como por ejemplo,
exoplanets.org\footnote{\url{http://exoplanets.org}}.

\paragraph{Objetivos:}

\begin{itemize}
  \item Introducir conceptos de estadística descriptiva a partir del análisis
    de variables dinámicas de los exoplanetas: distribuciones de masas y
    momentos (moda, media, mediana, varianza, desvíos, sesgo, y curtosis).
  \item Introducir conceptos básicos de programación estructurada, algoritmos y
    pseudo-código.
  \item Deducir y verificar las leyes de Kepler del Movimiento Planetario con
    los datos de los exoplanetas.
\end{itemize}

\paragraph{Fuentes a consultar:}

Además de las fuentes de datos en línea como
{\textit{Exoplanets}}\footnote{\url{http://www.exoplanets.org}}, el
{\textit{NASA Exoplanet
Archive}}\footnote{\url{http://exoplanetarchive.ipac.caltech.edu/}}, y el
{\textit{Open Exoplanet
Catalogue}}\footnote{\url{http://www.openexoplanetcatalogue.com/}}, cualquier
libro introductorio de probabilidad y estadística y el libro {\textit{Python
para todos}}\cite{Gonzalez-Duque2010}, disponible en
línea\footnote{\url{http://mundogeek.net/tutorial-python/}}. Los códigos y
datos se encuentran disponibles en el blog del curso y en el repositorio de
códigos\footnote{\url{https://github.com/asoreyh/IntroFisUIS}}.

\paragraph{Actividades a desarrollar:}

\begin{itemize}
  \item Interactuar con las bases de datos en línea, identificando los
    parámetros de interés para la realización de esta actividad: masa del
    planeta, masa de la estrella, parámetros orbitales (principalmente el
    semieje mayor y la excentricidad de la órbita). 
  \item Escribir en pseudo-código un algoritmo para calcular la la media y la
    desviación cuadrática media de los parámetros de estudio.
  \item Modificar un código en python simple para leer los datos de los
    exoplanetas y calcular la masa media y su desvío.
  \item Deducir y verificar la validez de las leyes de Kepler aplicando un
    código python sobre los datos obtenidos.
\end{itemize}

\subsection{Rocas navegantes del Valle de la Muerte}

En un artículo reciente se realizan medidas y proponen una posible solución al
misterio de las rocas andantes del Valle de la Muerte en California. La
pregunta de cuáles son los efectos que mueven las rocas de este valle desértico
se viene repitiendo por décadas\,\cite{Shelton1953} y las primeras medidas que
registran el movimiento de las rocas han sido tomadas por Norris y
colaboradores\cite{NorrisEtal2014}. Estos autores colocaron módulos de
geoposicionamiento satelital (GPS) en varias rocas y estudiaron su
desplazamiento y muestran que, aparentemente, las rocas se mueven sobre láminas
de hielo que se quiebran y deslizan.

\paragraph{Objetivos:}

\begin{itemize}
  \item Interpretar gráficas: velocidad \textit{vs} tiempo y posición
    \textit{vs} tiempo y, a partir de éstas obtener información sobre la
    dinámica del moviendo de las rocas
  \item Discutir los conceptos fuerzas de fricción en fluidos y entre
    materiales, haciendo énfasis en el significado de la fricción estática y
    dinámica y, como pueden ser calculadas a partir de los datos
  \item Discutir la nueva manera de comunicación de la ciencia en la cual los
    artículos están asociados a conjuntos de datos que pueden ser analizados    
\end{itemize}

\paragraph{Fuentes a consultar:}

Además del artículo de Norris y colaboradores\cite{NorrisEtal2014} con las
fuentes de datos asociadas, se consultarán algunos \textit{blogs}
especializados\footnote{
\url{http://www.nature.com/news/wandering-stones-of-death-valley-explained-1.15773}
y también \\ \url{http://www.racetrackplaya.org}} 

\paragraph{Actividades a desarrollar:}

\begin{enumerate}
  \item Algunas de las medidas surgidas en estos años han estimado que el
    coeficiente de fricción estático es $\mu_{e} \approx 0.15$. Estime la
    fuerza (¿máxima?) necesaria para iniciar y luego mantener en movimiento
    algunas de las rocas ``instrumentadas'' que aparecen en al Tabla 1del
    artículo de Norris y colaboradores. 
  \item El módulo de la fuerza de fricción en un fluido sobre un cuerpo que
    desplaza con una velocidad $v$ puede modelarse como $F_{f} = C \rho A
    v^2/2$, donde $C$ es un coeficiente de resistencia que depende de la forma
    del cuerpo, $\rho$ la densidad del fluido ($\rho 1.21$\,kg\,m$^{-3}$ en el
    Valle de la Muerte), y $A$ el área de la sección transversal que el cuerpo
    ofrece al fluido.
    \begin{enumerate}
      \item A partir de los datos de las figuras 5 y 9 (y los datos en la tabla
        suplementaria 1) del artículo de Norris y colaboradores estime el
        producto de constantes $C$ y $A$. Discuta sobre los posibles errores de
        esta estimación.
      \item Con los estimados anteriores haga un gráfico de la variación en el
        tiempo del coeficiente de fricción cinético para las rocas $A3$ y $A6$
        del experimento de Norris y colaboradores. ¿cuál es el coeficiente
        cinético medio en intervalos de $1h$, $4h$ y $8h$
    \end{enumerate}
\end{enumerate}

\subsection{Cambio Climático}\label{CambioClimatico}

Utilizando conceptos termodinámicos, en particular de temperatura y calor,
abordamos la discusión del cambio climático centrada en la generación de
dióxido de carbono por la actividad humana. Los datos muestran una clara
correlación entre el aumento de la quema de combustibles fósiles, el aumento de
la concentración de gases de invernadero en la atmósfera y el calentamiento
global, señalando que la especie humana es la mayor responsable del aumento en
el promedio de la temperatura\,\cite{Stocker2013}.

\paragraph{Objetivos}

Esta actividad tiene dos objetivos principales
\begin{itemize}
  \item utilizar conceptos estadísticos (media y desviación cuadrática media)
    para calcular variables físicas, discutir el significado de una magnitud
    física obtenida a partir de estos conceptos;

  \item estimular el uso de estimados gruesos (las llamadas estimaciones
    \textit{a la Fermi}), para obtener órdenes de magnitud de cantidades
    físicas relevantes de un problema.
\end{itemize}

\paragraph{Fuentes a consultar}

\begin{itemize}
  \item Indicadores del Banco
    Mundial\footnote{\url{http://data.worldbank.org/data-catalog/world-development-indicators?cid=GPD_WDI}} 
  \item Enerdata\footnote{\url{http://yearbook.enerdata.net/}}; 
  \item Intergovernmental Panel on Climate Change,
    IPCC\footnote{\url{http://www.ipcc.ch/}};
  \item Statistical Review of World Energy 2013,
    BP,\footnote{\url{http://www.bp.com/en/global/corporate/about-bp/energy-economics/statistical-review-of-world-energy.html}}
  \item Los códigos y datos se encuentran disponibles en el blog del curso y en
    el repositorio de
    códigos\footnote{\url{https://github.com/asoreyh/IntroFisUIS}}.
\end{itemize}

\paragraph{Actividades a desarrollar:}

\begin{itemize}
  \item Hacer un gráfico de la concentración de CO$_{2}$ como función del
    tiempo para todo el registro.
  \item Calcular el valor medio y el desvío muestral de las mediciones de la
    concentración de CO$_{2}$ del último millón de años, sin considerar el
    período reciente (antes de 1950)
  \item Calcular a cuantos ``sigmas'' de la media se encuentra el valor actual
    de la concentración.
  \item Calcular, haciendo una estimación \textit{a la Fermi}, cuantos
    kilogramos de CO$_{2}$  se liberaron en el año 2013.
\end{itemize}

\subsection{Fenómenos de Meteorología Espacial en los datos del proyecto LAGO}\label{SpaceWeather}

Es un fenómeno bien conocido que la actividad Solar modula el flujo de rayos
cósmicos de origen galáctico. Esta modulación en el flujo de rayos cósmicos
puede ser estudiada, por ejemplo, mediante el análisis de las modulaciones en
el flujo de partículas secundarias producidas durante la interacción de esos
rayos cósmicos con la atmósfera terrestre (ver por
ejemplo\,\cite{DassoAsorey2012, Asorey2013}). En esta experiencia, se presenta
a los estudiantes un conjunto preprocesado de datos del proyecto LAGO, para que
los estudiantes del curso de Introducción a la Física completen el análisis de
los mismos.

\paragraph{Objetivos}

Esta actividad tiene tres objetivos principales:

\begin{itemize}
  \item utilizar conceptos estadísticos (media y desviación cuadrática media)
    para calcular valores medios de los flujos de partículas a partir de los
    datos del proyecto LAGO;
  \item analizar y descorrelacionar el efecto de la presión atmosférica sobre
    el flujo de partículas al nivel de los detectores;
  \item encontrar en los datos analizados indicios de actividad solar como los
    llamados decrecimientos Forbush.
\end{itemize}

\paragraph{Fuentes a consultar}

Notas técnicas del proyecto LAGO relacionadas y artículos relacionados. Los códigos y datos se encuentran disponibles en el blog del curso y en el repositorio de códigos\footnote{\url{https://github.com/asoreyh/IntroFisUIS}}.

\paragraph{Actividades a desarrollar:}

\begin{itemize}
  \item Utilizar los códigos desarrollados para obtener una serie temporal con
    los promedios del flujo de partículas al nivel del detector. 
  \item Hacer un gráfico del flujo sin corregir como función del tiempo. 
  \item Hacer un gráfico del flujo observado como función de la presión
    atmosférica.
  \item A partir de una regresión lineal, obtener los parámetros de la
    anticorrelación entre el flujo observado y la presión atmosférica.
  \item Corregir los efectos de la presión atmosférica.
  \item Identificar señales características de actividad solar en los datos
    corregidos.
\end{itemize}

\subsection{Señales transitorias en los datos del proyecto LAGO}

Las fuentes de astropartículas se pueden dividir en periódicas y aperiódicas o
transitorias. Con la intensión de captar radiación de fenómenos transitorias
originados en destellos de rayos gamma, el proyecto LAGO tiene ubicado
detectores Cherenkov de agua en el monte Chacaltaya en Bolivia, a 5200\,m
s.n.m. 

\paragraph{Objetivos:}

\begin{itemize}
  \item Analizar los datos recolectados por detectores del proyecto LAGO para
    hallar señales de fuentes de astropartículas de tipo transitorio.
  \item Trabajar con grandes volúmenes de datos mediante acceso remoto vía
    protocolo ssh (\textit{Security Shell, por sus siglas en ingles}).
\end{itemize}

\paragraph{Fuentes a consultar:}

Trabajos realizados por miembros de la colaboración LAGO donde se describen los
datos recolectados por los detectores y el método de
análisis\,\cite{NunezQuinonezSarmiento2013,Sarmiento2012}.

\paragraph{Actividades a desarrollar:}

\begin{itemize}
  \item Realizar un informe donde se desglose la organización de los datos y el
    funcionamiento de cada una de las partes de uno de los detectores Cherenkov
    en Agua (WCD, por sus siglas en inglés) del proyecto LAGO.
  \item Obtener candidatos a fuentes de astropartículas usando el método de la
    ventana corrediza. Este método consiste en calcular el promedio y la
    desviación cuadrática en una ventana de datos y lo compara con el
    \textit{bin} central e inicia un corrimiento de un \textit{bin} hasta
    finalizar el archivo de datos. Los códigos usados para este análisis han
    sido realizados con ROOT\footnote{\url{https://root.cern.ch/}} y python.
  \item Elaborar un informe con la validación de los candidatos a fuentes
    hallados en el análisis anterior. Para esta actividad se usaran los datos
    de montaje y ubicación geográfica del experimento, así como las condiciones
    iniciales de puesta en funcionamiento del detector. A estos datos se les
    conoce como metadata.
\end{itemize}

\subsection{Nuevos datos LAGO y análisis de la línea base}

El proyecto LAGO ha diseñado y producido una nueva tarjeta electrónica que
optimiza el proceso de recolección de los datos, a la vez que permite almacenar
cada uno de los pulsos capturados por el detector cada vez que una partícula
ingresa al tanque. Basado en la experiencia con los datos recolectados con la
anterior electrónica se hace necesario analizar la línea base de los datos para
verificar la correcta recolección de los mismos y en caso de que esta falle
poder realizar los ajustes pertinentes.

\paragraph{Objetivos:}

\begin{itemize}
  \item Estudiar la organización de los nuevos datos.
  \item Hacer un \textit{script} en python que verifique el correcto
    funcionamiento de la línea base del detector.
\end{itemize}

\paragraph{Fuentes a consultar:}

Notas técnicas de la colaboración LAGO y manuales de python.

\paragraph{Actividades a desarrollar:}
\begin{itemize}
  \item Realizar un informe que describa la organización de los los datos y los
    metadatos recolectados por los WCD.
  \item Representar gráficamente el conteo de partículas contra el tiempo de
    todo el archivo.
  \item Escribir un código en python que calcula el promedio y la desviación
    cuadrática. Representar gráficamente el histograma de la distribución
    estadística en términos del exceso respecto a la desviación.
  \item Realizar un código que establezca un condicional para la línea base en
    un intervalo de aceptación. En caso de que alguno se encuentre fuera de
    este intervalo debe emitir un aviso de error.
\end{itemize}

\end{document}